\documentclass{PoS}

\usepackage{epsfig}
\usepackage{amsmath}

\def\mres{m_{\rm res}}
\newcommand{\MSbar}{\overline{\mbox{MS}}}
\newcommand{\msbar}{\overline{\mbox{\scriptsize MS}}}

\def\mom{\mbox{\scriptsize MOM}}
\def\smom{\mbox{\scriptsize SMOM}}

\def\simge{
    \mathrel{\rlap{\raise 0.511ex
        \hbox{$>$}}{\lower 0.511ex \hbox{$\sim$}}}}
\def\simle{
    \mathrel{\rlap{\raise 0.511ex
        \hbox{$<$}}{\lower 0.511ex \hbox{$\sim$}}}}

\def\cancel#1#2{\ooalign{$\hfil#1\mkern1mu/\hfil$\crcr$#1#2$}}
\def\slash#1{\mathpalette\cancel{#1}}

\title{Quark mass renormalization with non-exceptional momenta}

\ShortTitle{Quark mass renormalization with non-exceptional momenta}

\author{\speaker{Yasumichi Aoki}
         \\
        RIKEN BNL Research Center, Brookhaven National Laboratory,
	Upton, NY 11973, USA\\
        E-mail: \email{yaoki@bnl.gov}}
\author{RBC and UKQCD collaborations}

\abstract{Renormalization conditions imposed on quark bilinear vertex functions
in the conventional RI/MOM scheme use exceptional momentum configurations.
With practical values for the lattice cutoff, these vertex functions
are contaminated with unwanted low energy physics (pion pole, zero modes,
etc), which is a large source of systematic error.  These effects can be
reduced by using non-exceptional momenta. We discuss the quark mass
renormalization with non-exceptional momenta  using 2+1 flavor domain
wall fermions based on a recently proposed RI/SMOM scheme.}

\FullConference{The XXVI International Symposium on Lattice Field Theory\\
		 July 14-19 2008\\
		 Williamsburg, Virginia, USA}

\begin{document}

\section{Introduction}

Quark masses are fundamental parameters in QCD.
Recent lattice computations made it possible to calculate light 
quark masses up to strange consistently taking into account the light
sea quark effects.
The $2+1$ flavor domain wall fermion (DWF) calculation
reported \cite{Allton:2008pn},
\begin{eqnarray}
 m_{ud}^{\msbar}(2\mbox{GeV}) & = & 3.71(0.16)_{\mbox{stat}}
  (0.18)_{\mbox{syst}}(0.33)_{\mbox{ren}}\mbox{MeV},\\
 m_{s}^{\msbar}(2\mbox{GeV}) & = & 107.3(4.4)_{\mbox{stat}}(4.9)_{\mbox{syst}}
  (9.7)_{\mbox{ren}}\mbox{MeV}.
\end{eqnarray}
The first error is statistical.
The second error is the systematic uncertainty in the determination
of the bare quark mass in the lattice theory.
It is dominated by the 
discretization error and will be significantly reduced when 
the ensemble on a finer lattice is analyzed: 
we are currently generating these configurations 
which will enable us to perform a continuum extrapolation. 
In this paper we discuss the reduction of the third error which arises
in the renormalization of the mass. 
There are two dominant 
contributions to the error:
(i) The non-perturbative renormalization of the mass in the RI/MOM
scheme. We estimate the corresponding uncertainty to be about 7\%
due to the contamination by chiral symmetry breaking effects as
explained below.
(ii) The matching from the RI/MOM to the $\MSbar$ scheme.
The perturbative series for this matching is known to 3-loops
\cite{Chetyrkin:1999pq,Gracey:2003yr} but converges very poorly.
The uncertainty is estimated to be about 6\%.

As was demonstrated in Ref.~\cite{Aoki:2007xm}, the unwanted
non-perturbative contaminations in RI/MOM scheme due to spontaneous
chiral symmetry breaking could  be reduced 
by changing the scheme to one in which no exceptional momenta
are present. The argument stems from the Weinberg's theorem 
\cite{Weinberg:1959nj}
on the behavior of the vertex function for large external
momenta, where a set of external momenta which has zero
partial sum is called exceptional.

In this article after the construction of an RI/MOM scheme with
non-exceptional momenta for the quark bilinears is summarized,
the method is applied to a data set used in the conventional RI/MOM
renormalization with $2+1$ flavor DWFs
\cite{Aoki:2007xm}. The new results are compared with the conventional
RI/MOM results.

\section{RI/SMOM scheme for quark mass}

\subsection{Conventional RI/MOM scheme}

We start briefly summarize the original RI/MOM scheme.
A mass renormalization factor is completely fixed by introducing the
two renormalization conditions on the quark propagator.
For the conventional RI/MOM scheme, the conditions
on the Landau-gauge propagator read
\begin{eqnarray}
 \left. \frac{1}{12}\mbox{Tr}\left[-i\frac{\partial}
			      {\partial \slash{p}}S_R^{-1}(p)\right]
 \right|_{p^2=\mu^2} & = & 1,\label{eq:RIq}\\
 \lim_{m_R\to 0}\frac{1}{12 m_R} \mbox{Tr} [S^{-1}_R(p)]_{p^2=\mu^2}
  & = & 1,\label{eq:RIm}
\end{eqnarray}
which are imposed at the mass-less point.
The renormalized quark propagator and mass are related to the
bare ones through
\begin{align}
 S_R(p) = Z_q(\mu) S_B(p),&& m_R=Z_m(\mu) m_B.
\end{align}
Eq.~(\ref{eq:RIq}) determines $Z_q^{\mom}(\mu)$ which is needed for the
Eq.~(\ref{eq:RIm}), which in turn fixes $Z_m^{\mom}(\mu)$. 
In the continuum theory, RI/MOM scheme wave function renormalization
condition Eq.~(\ref{eq:RIq}) can be rewritten in terms of the 
renormalization condition on the bare amputated Green
function $\Pi$ of the vector current
through the Ward-Takahashi identity as
\begin{equation}
 \left. \frac{1}{Z_q^{\mom}} \frac{1}{48} \mbox{Tr} [\gamma_\mu \Pi_{V_\mu}
  (p)]\right|_{p^2=\mu^2} = 1,
 \label{eq:RIV}
\end{equation}
with $Z_V=1$.
A similar relation applies for the axial vector vertex function,
but with a contamination of a non-perturbative effect with $1/p^2$
suppression \cite{Martinelli:1994ty_short},
\begin{equation}
 \left. \frac{1}{Z_q^{\mom}} \left(\frac{1}{48} \mbox{Tr} [\gamma_5 \gamma_\mu \Pi_{A_\mu}
  (p)] + \frac{c_{NP}}{p^2}+\cdots\right)\right|_{p^2=\mu^2} = 1,
 \label{eq:RIA}
\end{equation}
with $Z_A=1$.
The momentum configuration must be stated to fix the renormalization
condition. For these to give equivalent renormalization condition as 
Eq.~(\ref{eq:RIq}), when the momentum $p$ comes in through one fermion line
the same $p$ must go out via the other fermion line. 
This is an exceptional momentum configuration ($p_1+p_2=p-p=0$). 
It is shown from the Weinberg's theorem \cite{Weinberg:1959nj} that the
difference of the 
vector and axial vector vertex amplitude is $\sim 1/p^2$ \cite{Aoki:2007xm}.
This is consistent with the existence of the $1/p^2$ contamination term
in Eq.~(\ref{eq:RIA}), which Martinelli {\it et al.} derived through
operator product expansion in Ref.~\cite{Martinelli:1994ty_short}. 

On the lattice with DWFs, the use of vector or
axial vector vertex function  (Eqs.~(\ref{eq:RIV}, \ref{eq:RIA})) 
has an advantage over the quark propagator (Eq.~(\ref{eq:RIq})) in calculating
the quark wave function renormalization.  The derivative with respect to
momentum is not practical on the lattice as the
momenta are quantized 
on the finite volume lattice. A similar scheme sometimes called as RI',
in which Eq.~(\ref{eq:RIq}) is replaced with 
\begin{equation}
  \left. \frac{1}{12p^2}\mbox{Tr}
   [-i \slash{p} S_R^{-1}(p)]
  \right|_{p^2=\mu^2} = 1,\label{eq:RI'q}
\end{equation}
is free from the derivative.
But, naive implementations of Eq.~(\ref{eq:RI'q}) on the lattice
introduce the tree revel $(pa)^2$ error, which is sizable at
the momentum range we use \cite{Blum:2001sr}.
DWFs can utilize the conserved
axial vector current and provide a precise estimate of $Z_A(=Z_V)$
\cite{Allton:2008pn}
of the local currents, which in turn allows one to use 
Eqs.~(\ref{eq:RIV}), (\ref{eq:RIA}) with a correction of a $Z_A$ factor to
get $Z_q^{\mom}$. 

By similar reasons, the mass renormalization should be 
calculated through bilinear operator renormalization using the relation
$Z_m=1/Z_S=1/Z_P$.
In principle, the scalar and pseudoscalar renormalization factors 
can be determined at large momenta by imposing the conditions
\begin{align}
  \left. \frac{Z_S}{Z_q} \frac{1}{12} \mbox{Tr} [\Pi_S
  (p)]\right|_{p^2=\mu^2} = 1, &&
  \left. \frac{Z_P}{Z_q} \frac{1}{12} \mbox{Tr} [\gamma_5 \Pi_P
  (p)]\right|_{p^2=\mu^2} = 1.
 \label{eq:RISP}
\end{align}
At finite momenta however, $Z_S$ may differ from $Z_P$
due to spontaneous chiral symmetry breaking.
In particular, for the exceptional momentum case, 
one needs to subtract the pion pole (for $P$) or
double pole (for $S$ with quenching) to remove the divergence
of $Z_P$ (and $Z_S$ quench) in the chiral limit.

\subsection{RI/SMOM scheme}
Non-exceptional momenta do not have zero partial sum,
which suppresses, in the Feynman diagram, the small momentum flow leading to
non-perturbative contamination. Among various choices of non-exceptional
momenta, we adopt the symmetric one $\mu^2=p_1^2=p_2^2=q^2$ where $q=p_1-p_2$.
This choice is convenient because only one invariant is involved
($2p_1\cdot p_2$ is also $\mu^2$).
Here we briefly review the SMOM (symmetric MOM) scheme mass renormalization
which makes use of the symmetric momentum configuration.
The SMOM scheme is discussed in detail in Ref.~\cite{Sturm:SMOM}.
We will demonstrate in the next section 
how the use of this renormalization scheme reduces the
unwanted non-perturbative contaminations compared to the conventional
MOM scheme.

Other than changing the momentum configuration
the SMOM scheme follows the same steps as MOM scheme. The
renormalization conditions are defined using trace conditions with 
specified projectors on the vertex functions $\Pi_O$.
For the scalar and pseudoscalar operators, the same projection operators
as MOM scheme $\frac{1}{12}1$, $\frac{1}{12}\gamma_5$ as shown in
Eq.~(\ref{eq:RISP}) are used.  The vector and
axialvector operator will be used to calculate $Z_q$.  Original MOM
scheme uses $\frac{1}{48}\gamma_\mu$, $\frac{1}{48}\gamma_5\gamma_\mu$,
by which one 
can relate the traced vertex function to $Z_q^{\mom}$
Eq.~(\ref{eq:RIq}). 
If we used these projection operators for the symmetric (non-exceptional)
momenta, the resulting $Z_q$ would completely differ from that of MOM
scheme (or RI' scheme).  Instead, we adopt $\frac{1}{12q^2}\slash{q}q_\mu$,
$\frac{1}{12q^2}\gamma_5\slash{q}q_\mu$. One can show that the use
of these projection operators give $Z_q^{\mbox{RI'}}$ through the vector
and axial vector Ward-Takahashi identities.  The matching of $Z_q$
of RI' and $\MSbar$ has been calculated to three loops
\cite{Chetyrkin:1999pq,Gracey:2003yr}, which one can just use or can use for the
check against the calculation with the vertex functions 
of vector and axialvector in the SMOM scheme.
It is worth mentioning that RI' and RI/MOM $Z_q$ are
same up to one loop.  Thus, the resulting $Z_q$ from the vector
and axialvector current of SMOM scheme should be close to that
of original MOM scheme.
The perturbative matching of quark mass from SMOM to $\MSbar$, 
$m^{\msbar}(\mu)=C_m(\smom\to\msbar)\cdot m^{\smom}(\mu)$
has been calculated to one loop \cite{Sturm:SMOM} as
\begin{equation}
 C_{m}(\smom\to\msbar) = 
  1 - \frac{\alpha_s}{4\pi} C_F\times (0.484-0.172\xi) + O(\alpha_s^2),
  \label{eq:SMOM-MSbar}
\end{equation}
where $\xi$ is the gauge parameter.
The same quantity for the original RI/MOM to $\MSbar$ has much larger
correction (both constant and linear coefficient of $\xi$):
\begin{equation}
 C_{m}(\mom\to\msbar) = 
  1 - \frac{\alpha_s}{4\pi} C_F\times (4-\xi) + O(\alpha_s^2).
\end{equation}
The one loop correction for the Landau gauge ($\xi=0$) is 1.5\% for SMOM and
12\% for MOM at $\mu=2$ GeV. The three loop correction is still large:
6\% for MOM, which was taken as a conservative estimate of systematic error
of perturbative matching \cite{Aoki:2007xm}.
The small correction of SMOM scheme is realized through
cancellation of finite terms depending on the
momentum structure. We have not understood if this is an universal
property with SMOM scheme, which would persists beyond one loop.

\section{Numerical test of the RI/SMOM scheme}

We test the RI/SMOM scheme using the $N_f=2+1$ DWF data set
\cite{Aoki:2007xm} at $a^{-1}\simeq 1.7$ GeV on $16^3\times 32$ lattice
with $L_s=16$, $M_5=1.8$. The quark propagators have been calculated
with the point source \footnote{The statistical error could be much
improved if the volume source was used 
\cite{Brommel:pos_lat08,Kelly:pos_lat08,Wennekers:pos_lat08}.}.

\begin{figure}
 \begin{center}
  \epsfig{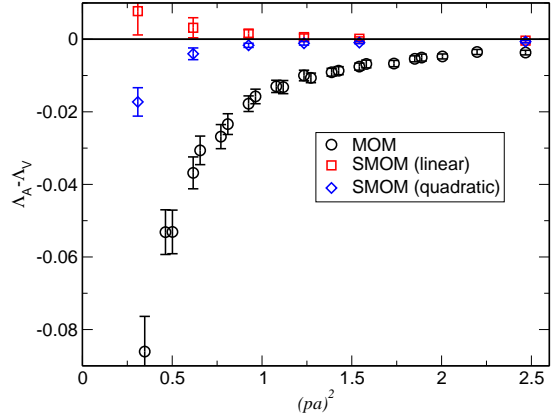}
  \caption{$\Lambda_A-\Lambda_V$ 
  in the chiral limit as function of 
  $p^2$ for MOM and SMOM scheme, where results with linear and 
  quadratic chiral extrapolations in quark mass are shown.}
 \label{fig:A-V}
\end{center}
\end{figure}

Let us first look at the difference of the vector and axial vector
vertex amplitude,
\begin{align}
 \Lambda_V^{\mom}=\frac{1}{48} \mbox{Tr} [\gamma_\mu \Pi_{V_\mu}],
 &&
 \Lambda_A^{\mom}=\frac{1}{48} \mbox{Tr} [\gamma_5\gamma_\mu \Pi_{A_\mu}]
 \label{eq:LRIV}
\end{align}
for the exceptional momentum and for the symmetric (non-exceptional)
momentum with
\begin{align}
 \Lambda_V^{\smom}=\frac{1}{12q^2} \mbox{Tr} [\slash{q}q_\mu \Pi_{V_\mu}],
  &&
 \Lambda_A^{\smom}=\frac{1}{12q^2} \mbox{Tr} [\gamma_5\slash{q}q_\mu \Pi_{A_\mu}].
 \label{eq:sRIV}
\end{align}
Fig.~\ref{fig:A-V} shows the differences in the chiral limit as
functions of $p^2$ ($=q^2$).
The original MOM scheme has non-zero difference due to the spontaneous
chiral symmetry breaking. 
The decrease of the difference as momentum increases is 
due to the recovery of the symmetry and the size is up to 1\% of
the average $(\Lambda_A+\Lambda_V)/2$ in the region of the momentum we
use $(pa)^2> 1.3$.
The difference is much suppressed
for the SMOM scheme.  The linear chiral extrapolation gives results
consistent with zero. The quadratic extrapolation gives non-zero value,
but one order of magnitude smaller than MOM.

\begin{figure}
 \begin{center}
  \epsfig{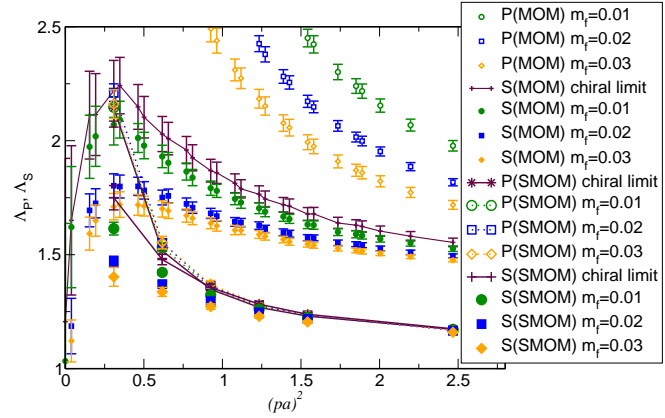}
  \caption{$\Lambda_S$ and $\Lambda_P$ for MOM and SMOM
  scheme. $m_f$ refers to the average $u$, $d$ mass.
  Points connected with the solid lines show the values in the 
  two-flavor unitary chiral limit $(m_f+\mres\to 0)$.
  }
 \label{fig:SP}
\end{center}
\end{figure}

The scalar and pseudoscalar vertex amplitudes with MOM and SMOM scheme
are shown in Fig.~\ref{fig:SP}.
A large difference between $P$ and $S$ is observed for the MOM scheme.
Since $\Lambda_P$ diverges as $\sim 1/m$ in the chiral limit due to the
existence of pion pole
$\langle\overline{\psi}\psi\rangle/m_\pi^2$
and $\Lambda_S$ stays finite, the difference 
becomes infinite. One should note that these quantities are identical
to all order in perturbation theory. The symmetry is badly broken
for the non-perturbative renormalization.
On the other hand, $\Lambda_P$ and $\Lambda_S$ are consistent with
each other for the SMOM case at larger momentum $(pa)^2\simge 1$.

As our gauge ensembles have been sampled at single value of
nearly physical strange mass, we have a systematic error
from $m_s\ne 0$ even after the two-flavor unitary chiral extrapolation.
This error in the MOM scheme, which turned out to be 7\%, was
estimated from the response of the $\Lambda_S$ to the $u$, $d$ quark mass
in Ref.~\cite{Aoki:2007xm}.  This error may represent the tolerance
of this particular quantity to the emergence of the low energy scale
$(\sim \Lambda_{QCD})$. So even if the $m_s\to 0$ limit was performed, error 
of similar size would remain due to the non-perturbative effect whose energy
scale is about the same. Now, in the SMOM case, as shown in the figure,
mass dependence is greatly reduced. If we adopt the same method,
the systematic error of SMOM is about $3$\% for the scalar
or negligible (comparable to the statistical error) for the pseudoscalar.

\begin{figure}
 \begin{center}
  \epsfig{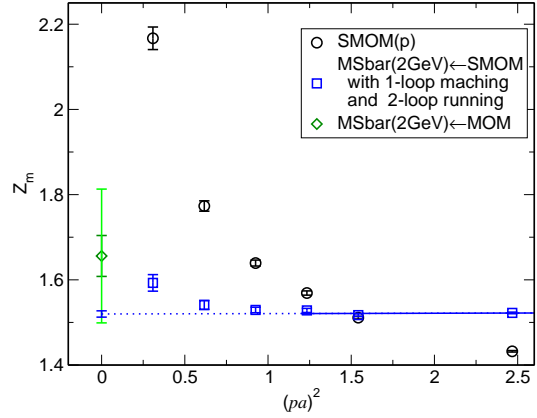}
  \caption{Preliminary results of $Z_m^{\smom}(p)$ as a function of
  renormalization scale $p$ 
  and $Z_m^{\msbar}(\mu=2\mbox{GeV})$ as a function of SMOM$\to\MSbar$
  matching scale $p$. The extrapolation $(pa)^2\to 0$ is shown 
  with statistical error only, which is consistent with the
  $Z_m^{\msbar}(\mu=2\mbox{GeV})$ through MOM scheme with
  the large systematic error.}
 \label{fig:Z_m}
\end{center}
\end{figure}
Black symbols in Fig.~\ref{fig:Z_m} show preliminary results
of $Z_{m}=(\Lambda_S+\Lambda_P)/\{Z_A(\Lambda_A+\Lambda_V)\}$ with SMOM
scheme (in the chiral limit) as a function of renormalization scale $p$,
where the $Z_A$ \cite{Allton:2008pn} estimated form hadronic two point
functions is used. 
Matching to $\MSbar$ with Eq.~(\ref{eq:SMOM-MSbar}) and running
to $\mu=2$ GeV with the two-loop anomalous dimension, one obtains
the blue symbols as a function of matching scale $p$. 
The extrapolation
$(pa)^2\to 0$ using the points $(pa)^2\simge 1.2$ gives a consistent
result with the same quantity but through original MOM scheme
\cite{Aoki:2007xm} with 3-loop matching and 4-loop running.

\section{Conclusion}

The RI/SMOM scheme, constructed in the framework of the conventional
RI/MOM scheme with the use of non-exceptional momenta,
works very well for reducing non-perturbative
contamination for the quark mass renormalization. The systematic error
is reduced to 
3\%
level for $Z_m$, while it was 7\% 
for the original MOM scheme. This shows the success of the
SMOM scheme which was designed to reduce the unwanted non-perturbative
contamination.
Another systematic error is from truncation in the
perturbative matching to $\MSbar$.
If we estimate the systematic error for the SMOM scheme
from the size of $O(\alpha_s)$ at our typical momentum size $\mu=2$
GeV, it is 1.5 \%, which is much smaller than 6 \% at
$O(\alpha_s^3)$ for the MOM scheme.
Further discussions are needed for the better
understanding of the systematic error of the perturbative matching.

The first non-trivial test of the SMOM scheme was successful.
Application to other bilinear operators such as tensors
would be straightforward. Similar scheme can be constructed for
four-quark operators for $K^0-\overline{K^0}$ mixing in the standard
model and beyond, and for $K\to\pi\pi$ decays.

\vspace{18pt}
We thank Christian Sturm for collaborating on the RI/SMOM renormalization.
Presented numerical data are obtained through reanalyzing
the published data computed on the QCDOC machines at RIKEN BNL Research
Center,  Columbia University and University of Edinburgh.

%

\bibliography{paper}

\end{document}